\documentclass[lettersize,journal]{IEEEtran}
\usepackage{amsmath,amsfonts}
\usepackage{algorithmic}
\usepackage{algorithm}
\usepackage{array}
\usepackage[caption=false,font=normalsize,labelfont=sf,textfont=sf]{subfig}
\usepackage{textcomp}
\usepackage{stfloats}
\usepackage{url}
\usepackage{multirow}
\usepackage{tabularx,booktabs} 
\usepackage{verbatim}
\usepackage{graphicx}
\usepackage{cite}
\usepackage{makecell}
\usepackage{url}
\usepackage{relsize} 
\usepackage{color}
\makeatletter
\let\NAT@parse\undefined
\makeatother
\usepackage{hyperref}  

\definecolor{b}{rgb}{0,0,0} 

\begin{document}

\title{Toward Sustainable Subterranean mMTC: Space-Air-Ground-Underground Networks Powered by LoRaWAN and Wireless Energy Transfer}

\author{Kaiqiang Lin 
        and Mohamed-Slim Alouini
        
\thanks{This work was supported in part by the ENERCOMP consortium jointly funded by King Abdullah University of Science and Technology. (\textit{Corresponding author: Kaiqiang Lin})}      

\thanks{Kaiqiang Lin and Mohamed-Slim Alouini are with King Abdullah University of Science and Technology, Kingdom of Saudi Arabia.}\vspace{-5mm}
}


\maketitle

\begin{abstract}
Wireless underground sensor networks (WUSNs), which enable real-time sensing and monitoring of underground resources by underground devices (UDs), hold great promise for delivering substantial social and economic benefits across various verticals. However, due to the harsh subterranean environment, scarce network resources, and restricted communication coverage, WUSNs face significant challenges in supporting sustainable massive machine-type communications (mMTC), particularly in remote, disaster-stricken, and hard-to-reach areas. To complement this, we conceptualize in this study a novel space-air-ground-underground integrated network (SAGUIN) architecture that seamlessly incorporates satellite systems, aerial platforms, terrestrial networks, and underground communications. On this basis, we integrate LoRaWAN and wireless energy transfer (WET) technologies into SAGUIN to enable sustainable subterranean mMTC. We begin by reviewing the relevant technical background and presenting the architecture and implementation challenges of SAGUIN. Then, we employ simulations to model a remote underground pipeline monitoring scenario to evaluate the feasibility and performance of SAGUIN based on LoRaWAN and WET technologies, focusing on the effects of parameters such as underground conditions, time allocation, LoRaWAN spread factor (SF) configurations, reporting periods, and harvested energy levels. Our results evidence that the proposed SAGUIN system, when combined with the derived time allocation strategy and an appropriate SF, can effectively extend the operational lifetime of UDs, thereby facilitating sustainable subterranean mMTC. Finally, we pinpoint key challenges and future research directions for SAGUIN.
\end{abstract}

\section{Introduction}
Space-air-ground integrated networks (SAGIN) combine space, aerial, and terrestrial communication technologies to deliver seamless global connectivity~\cite{SAGReview}. To date, existing SAGIN architectures primarily focus on communication and sensing systems operating above the Earth's surface, i.e., in terrestrial and maritime domains, while neglecting underground soil environments. Notably, underground monitoring is essential for many applications including but not limited to:

\begin{itemize}
    \item \textbf{Agriculture:} By 2050, the global population is projected to reach $9.7$ billion, doubling the demand for food. However, traditional agriculture’s inefficiency, resource waste, and poor adaptability to climate and soil variability hinder its ability to meet this growing need. This calls for smart agriculture to enable data-driven resource management. For instance, in Saudi Arabia, agricultural water usage was reduced by $50\% $, crop yields were boosted by $300\%$, and fertilizer use was cut by $40\%$ through real-time underground soil sensing and timely decision-making~\cite{AriSupport}.

    \item \textbf{Underground Pipeline Transportation:} Underground pipelines serve as the lifelines of modern civilization by supplying water, natural gas, and crude oil over hundreds to thousands of kilometers. However, from 2010 to 2021, the PHMSA recorded over $1,200$ underground oil and gas pipeline incidents in the United States, resulting in substantial economic losses, severe environmental damage, and human casualties~\cite{PipeSupport}. This underscores the urgent need for real-time underground monitoring systems capable of early warning, rapid leak detection, and precise fault localization to enhance emergency response and strengthen infrastructure resilience.
    
    \item \textbf{Post-disaster Rescue:} According to the United States Geological Survey, over $200$ earthquakes with magnitudes exceeding $6.5$ have occurred since 2020~\cite{EqSupport}. The first $72$ hours following an earthquake are critical for maximizing survival rates. However, traditional rescue systems still rely heavily on manual operations, which limits their coverage, accuracy, and efficiency. To address these limitations, underground sensing technologies offer real-time monitoring and accurate localization of trapped individuals, significantly improving the speed and effectiveness of post-disaster rescue operations.
\end{itemize}

We have recently demonstrated the feasibility of integrating low-power wide-area network (LPWAN)-grade massive machine-type communication (mMTC) wireless technologies with wireless underground sensor networks (WUSNs) for real-time underground monitoring~\cite{LinLoRaWUSNs}. However, many of the aforementioned applications are envisioned to operate in remote or disaster-affected areas, where communication infrastructure may not be readily available. Due to scarce network resources, severe attenuation in underground soil, and economic constraints, mMTC-based WUSNs cannot provide real-time monitoring of large-scale underground infrastructure in such regions~\cite{LinMag}.

To overcome this critical limitation and ensure resilient communication services, we propose extending the SAGIN architecture into the subterranean domain, introducing a novel space-air-ground-underground integrated network (SAGUIN) paradigm. Within this framework, the integration of LoRaWAN and wireless energy transfer (WET) technologies is leveraged to enable sustainable subterranean mMTC. To the best of the authors’ knowledge, this is the first work to assess the feasibility of a SAGUIN system powered by LoRaWAN and WET technologies in prolonging the operational lifetime of UDs. Our main contributions are summarized as follow:
\begin{enumerate}
    \item We offer an overview of the technical background for sustainable subterranean mMTC, conceptualize a general SAGUIN architecture in both communications and energy aspects, and discuss its potential use cases and implementation challenges. Compared with the traditional wireless-powered underground sensor network (WPUSN) solution, our proposed SAGUIN system enables UDs to harvest energy from both aboveground power beacons and unmanned aerial vehicles (UAVs), integrates WET and LoRaWAN technologies, and deploys gateways on non-terrestrial network (NTN) platforms, to achieve sustainable and large-scale underground monitoring in hard-to-reach regions.

    \item We develop a simulation model featuring the realistic remote underground pipeline monitoring scenario and gain sight into the effects of underground conditions, LoRaWAN spread factor (SF) configurations, and WET duration on the performance of the SAGUIN system. The presented results highlight the interplay between network reliability and energy efficiency, which inspires the subsequent resource allocation optimization.
    
    \item We jointly design the LoRaWAN SF configuration and the time allocation between WET and data transmission with the aim of system lifetime maximization. Simulation results reveal that an optimal time allocation combined with an appropriate SF configuration can substantially improve the operational lifetime of UDs, where its performance is inherently influenced by the number of UDs, reporting periods, and harvested energy levels. Furthermore, we identify potential research directions for practical SAGUIN deployment.
\end{enumerate}

\section{Background and Technical Development}
\subsection{WUSNs and LPWAN-grade mMTC Technologies}
WUSNs do not require high data rates and can tolerate relatively high latency; however, energy‐efficient communication is essential for the long‐term operation of battery‐powered UDs. Consequently, LPWAN technologies, designed for long-range, low-power, low-data-rate, and low-cost mMTC applications, are therefore well suited for WUSNs. Nowadays, LoRaWAN, NB-IoT, and Sigfox are the leading technologies in the LPWAN market. Among them, NB-IoT and Sigfox are operated by dedicated mobile network operators, limiting their deployment flexibility and often making them infeasible in remote or post-disaster areas. In contrast, LoRaWAN, which employs Semtech’s patented LoRa physical layer, allows users to deploy and manage private networks independently of any specific operator, offering greater flexibility~\cite{LoRaWANreview}. Therefore, in this study, we consider LoRaWAN a cost-effective wireless communication solution due to its broader coverage, higher energy efficiency, lower hardware costs, flexible deployment, and freedom from traffic subscription fees.

Recent experimental studies have shown that underground‑to‑aboveground communication for LoRaWAN devices can achieve $50$~m at a burial depth of $0.4$~m~\cite{LinLoRaWUSNs}. Nevertheless, the communication range is constrained by substantial signal attenuation in subterranean channels, which is affected by soil texture, bulk density, volumetric water content (VWC), burial depth, and UDs' operating frequency, etc. Furthermore, severe underground attenuation, scarce network infrastructure, and economic considerations pose significant challenges for large‑scale LoRaWAN‑based WUSN deployments in remote and post‑disaster areas. To overcome this limitation, since dense terrestrial gateway deployments are often economically impractical, one promising solution is to deploy gateways on NTN platforms. 

\subsection{Underground-to-Satellite Connectivity}
Satellite systems are typically categorized by orbital altitude into low Earth orbit (LEO, $500\sim2000$~km), medium Earth orbit (MEO, $2000\sim35786$~km), and geostationary Earth orbit (GEO, at $35786$~km). Compared with MEO and GEO, LEO satellites offer smaller coverage areas but benefit from lower launch costs, reduced signal attenuation, and minimal propagation delays, making them particularly well suited for internet of things (IoT) applications. With the expansion of satellite‐enabled connectivity services, integrating LoRaWAN technology with LEO satellite communications has emerged as a promising solution for massive IoT deployments. As of 2025, over $100$ LEO satellites employing LoRa modulation have been launched and are operational, collectively broadcasting more than $36$ million LoRa packets to ground stations~\cite{TinyGs}. Furthermore, recent studies have demonstrated the feasibility of LoRaWAN direct-to-satellite (DtS) connectivity for various use cases, including offshore wind monitoring, maritime and cargo tracking, and smart mining~\cite{AsadDtSReview}.

To expand the coverage of WUSNs and enable large-scale underground monitoring, the authors in~\cite{LinMag} extended the DtS approach to the subterranean domain and proposed the underground DtS (U-DtS) architecture featuring LoRaWAN technology. Through extensive modeling of a realistic farming scenario, the results reveal that the LoRaWAN technology can establish the reliable connectivity between massive UDs buried at a depth of some dozen centimeters and a LEO satellite. However, this work did not consider other NTN platforms, such as high-altitude platforms (HAPs) and UAVs, in subterranean domains for meeting the diverse requirements of communication services. Moreover, the study in~\cite{LinDRL} developed an optimal SF allocation scheme using a deep reinforcement learning approach to reduce co-SF interference and enhance system energy efficiency. Despite this optimization algorithm helps extend the lifetime of UDs, sustainable underground monitoring still requires a continuous external energy supply. 

\subsection{Sustainable Underground Monitoring}
Although the ultimate lifespan of UDs is constrained by hardware degradation, battery depletion is often the main cause of premature failure. As battery replacement requires labor-intensive, costly, and hardware-risking excavation, WUSNs are designed for long maintenance-free operation but still face three challenges: (i) limited battery capacity; (ii) high energy demand to overcome underground attenuation; and (iii) the impracticality of battery replacement. To address these challenges, three primary energy harvesting techniques are employed in the subterranean domain: magnetic resonant coupling, vibration harvesting, and radio-frequency (RF) WET approaches~\cite{UWETReview}. \textcolor{b}{Due to its longer transmission range, flexible deployment via UAVs, and capability to deliver reliable and continuous power without the strict alignment or proximity requirements of magnetic resonant coupling and vibration approaches, RF WET has been introduced as a promising energy harvesting solution, enabling sustainable underground monitoring and giving rise to WPUSNs~\cite{LiuWPUSNs}.} In a typical WPUSN system, UDs harvest RF energy from a dedicated power source via WET technology, which is then used for data transmission. Moreover, to improve WET efficiency in harsh underground environments, multi-antenna techniques, backscatter communication, channel state information (CSI)-free WET schemes, and reconfigurable intelligent surfaces (RIS) have been incorporated into WPUSNs~\cite{LinRIS}. 


Despite significant technical advancements in both WPUSN and U-DtS systems, it remains unclear how WET and LoRaWAN-based data transmission can be harmonized and integrated within the SAGUIN framework, as well as what factors influence its performance. To answer these questions, in the next section, we first present the proposed SAGUIN architecture enabled by LoRaWAN and WET technologies, followed by generating numeric results to assess its feasibility.

\section{Proposed SAGUIN Architecture and Implementation}
\begin{figure}[!t]
\centering
\includegraphics[width=3.45in]{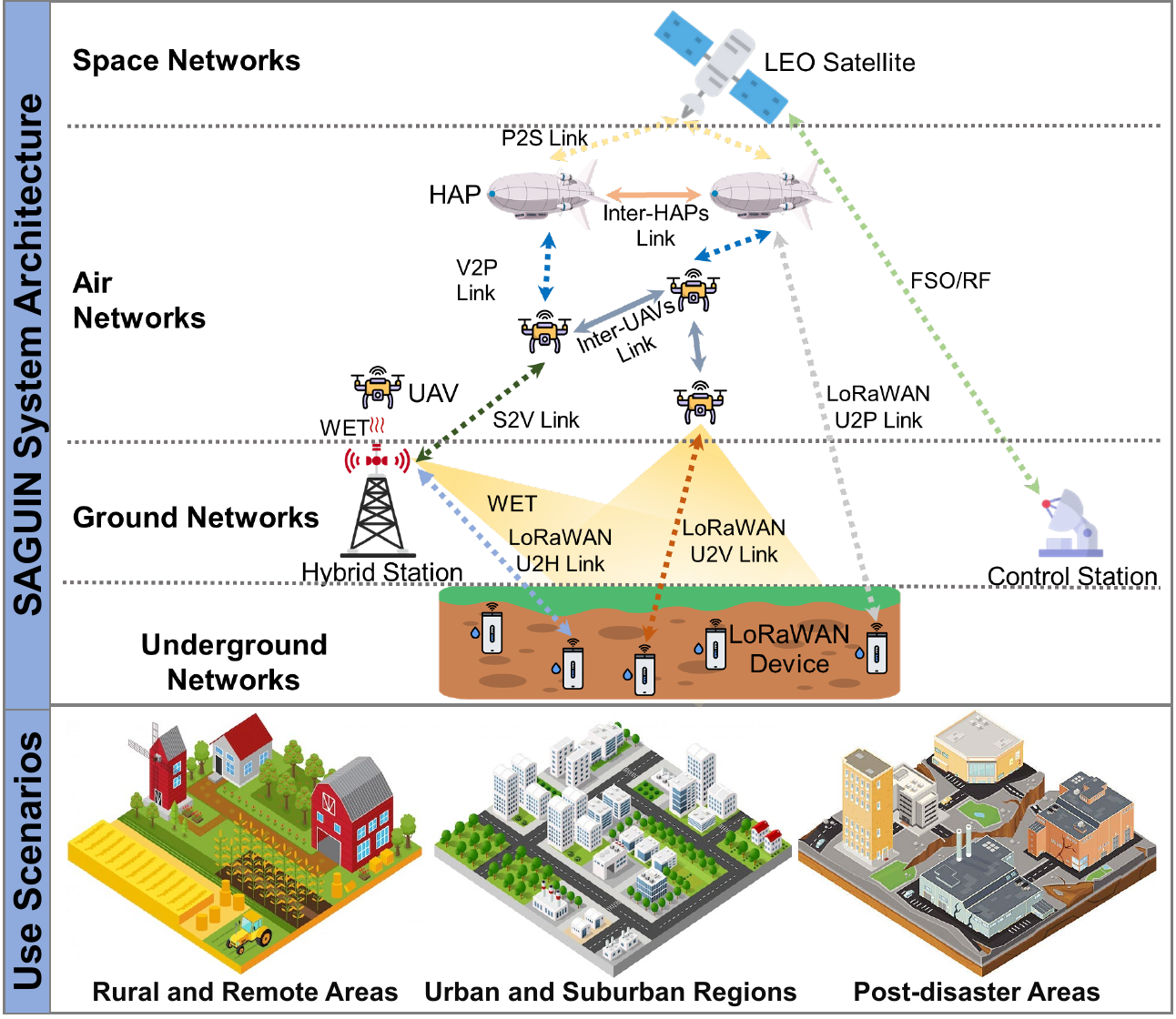}
\caption{Proposed SAGUIN system integrating LoRaWAN and WET solutions, where each UD harvests energy from aboveground hybrid stations and UAVs, and then utilizes this energy to communicate with various NTN platforms, including UAVs, HAPs, and LEO satellites. This system is applicable in urban, rural, and remote areas, as well as post-disaster regions, enabling various applications such as smart agriculture, underground pipeline monitoring, and post-disaster positioning.}
\label{fig_1}
\vspace{-3mm}
\end{figure}

In this section, we propose a SAGUIN system comprising space, air, ground, and underground components, as depicted in Fig,~\ref{fig_1}. By integrating LEO satellites with NTN platforms, terrestrial infrastructures, and UDs, the SAGUIN system attains a heterogeneous yet harmonized architecture capable of sustaining uninterrupted communication across urban, rural, remote, and post-disaster scenarios, thereby enhancing both operational flexibility and system resilience. The different network layers are interconnected via UD-to-hybrid station (U2H), UD-to-UAV (U2V), UD-to-HAP (U2P), hybrid station-to-UAV (H2V), UAV-to-HAP (V2P), and HAP-to-satellite (P2S) links. Furthermore, we advocate the integration of LoRaWAN and WET within the SAGUIN system to enable communication and energy harvesting functions for facilitating sustainable subterranean mMTC applications.

In a typical SAGUIN system, UDs harvest energy from both aboveground hybrid stations and UAVs via WET and utilize the harvested energy to upload data to the network server through LoRaWAN gateways. In urban environments with mature network infrastructure and a stable power grid, UDs connect directly to terrestrial LoRaWAN gateways with reliable backhaul. In remote and post-disaster scenarios, where dense terrestrial deployments are impractical or compromised, UDs connect to LoRaWAN gateways deployed on LEO satellites to extend coverage and enable subterranean mMTC, with UAVs and HAPs serving as relay stations to enhance communication reliability. LEO satellites then forward the aggregated sensor data to ground control stations via RF or free-space optical (FSO) feeder links.

\subsection{Communication Enablers}
\subsubsection{Space-Based Networks}
Space-based networks can operate independently of terrestrial networks while delivering reliable and resilient mission‑critical connectivity across all network layers. LEO satellites play a crucial role in subterranean mMTC applications, particularly in scenarios such as remote farming, underground pipeline monitoring in rural or remote areas, and post-earthquake rescue operations, due to their low latency, wide coverage, resilience, and ability to provide reliable connectivity to remote or disaster-stricken regions. By exploiting inter‑satellite mesh links and dedicated backhaul channels, LEO constellations can aggregate sensor data from UDs via ground base stations or NTN platforms, and forward it terrestrial control stations via RF and FSO communication for further processing and decision‑making.

\subsubsection{Air-Based Networks}
Air-based networks are a critical component of the SAGUIN system, as they can directly communicate with LEO satellites, ground networks, and UDs, helping to compensate for limitations in other network segments. The air-based network consists of various NTN platforms, such as HAPs and UAVs. 

HAPs typically refer to airborne systems positioned in the stratosphere above conventional aviation altitudes, and can be in the form of balloons or airships. They provide wide-area coverage and energy autonomy via solar power, making them a viable alternative for damaged or unavailable terrestrial base stations in remote or disaster-affected regions. Their quasi-stationary position and ability to establish direct line-of-sight links enable stable and continuous connectivity without the need for intermediate relays, offering advantages over LEO satellites in terms of service continuity. Moreover, HAPs can be rapidly deployed, are more resilient to adverse weather conditions than UAVs.

UAVs operate at lower altitudes (typically a few hundred meters) and offer flexible and on-demand wireless connectivity, particularly in scenarios where fixed terrestrial network infrastructure is unavailable, such as in remote regions or during natural disasters. They can act as temporary communication infrastructure to gather emergency information from individuals trapped in disaster-stricken areas. Note that the high energy demands of continuous propulsion and vulnerability to adverse weather conditions limit the suitability of UAVs for long-term and reliable underground monitoring.

\subsubsection{Ground-Based Networks}
In the SAGUIN system, ground-based networks typically include cellular networks, terrestrial Internet (which relies on physical cables such as fiber optics or copper wires to transmit data over long distances), ad-hoc networks, and power beacons. Unlike air-based networks, ground-based networks have relatively stable topologies, allowing them to provide reliable connectivity for UDs. In remote or post-disaster areas, ground networks may become partially or fully inoperative due to physical damage or limited coverage, highlighting the importance of integrating complementary air- and space-based solutions for enhancing network resilience.

\subsubsection{Underground-Based Networks}
Underground-based networks primarily consist of UDs equipped with sensing, communication, charging, and power supply modules. In typical subterranean mMTC applications, UDs protected by waterproof and corrosion-resistant enclosures are buried near monitoring targets to monitor key metrics of underground entities. Due to the limited range of underground-to-underground communication, a centralized network topology is typically adopted, allowing UDs to communicate directly with terrestrial gateways or NTN platforms, thereby reducing power-intensive underground multi-hop routing. From the perspective of cost, reliable connectivity, and energy efficiency, LoRaWAN technology is preferred for underground-based networks. In urban scenarios with mature infrastructure and a stable power grid, UDs connect directly to terrestrial LoRaWAN gateways, whereas in remote or post-disaster areas where dense terrestrial deployments are impractical or compromised, they connect to gateways deployed on NTN platforms to extend coverage and support large-scale underground monitoring. 

\subsection{Energy Harvesting Enablers}
Since the feasibility and sustainability of the proposed SAGUIN architecture hinge on energy availability at each network layer, we summarize platform-specific power supplies as follows. LEO satellites rely on solar panels for primary power generation, with rechargeable batteries sustaining operations during eclipse periods. HAPs similarly use large solar arrays during daylight, supplemented by onboard batteries for nighttime or low-sunlight operation. UAVs are powered by lithium-ion batteries, which can be charged from aboveground power beacons through wired or wireless approaches. The power supply for aboveground hybrid stations varies by scenario: remote sites use off-grid solar-plus-battery systems for self-sufficiency; urban sites draw from the grid to minimize costs; and post-disaster sites employ mobile diesel or portable-solar hybrids to ensure availability when infrastructure is compromised. \textcolor{b}{UDs are powered by waterproof and anti-corrosion treated lithium-ion batteries, can be recharged through techniques such as magnetic induction, vibration-based harvesting, and WET, among which WET is particularly promising in the proposed SAGUIN system due to its ability to provide reliable and continuous power without the alignment constraints of other approaches.}



\subsection{LoRaWAN and WET Integration Approach}
The integration of LoRaWAN and WET is introduced in the proposed SAGUIN architecture to enable reliable communication and sustainable energy harvesting. The implementation guidelines are details as follow. Each UD is equipped with a LoRaWAN radio module, a lithium-ion battery, an energy-harvesting module and sensors tailored to specific underground monitoring scenarios. During each reporting period, each UD first harvests energy from both aboveground power beacons and UAVs via WET during the WET phase. In the subsequent data transmission phase, each UD utilizes the harvested energy to collect sensor information, selects an appropriate SF configuration, and then uploads the data to the network server via LoRaWAN gateways deployed on NTN platforms.

\textcolor{b}{Although WET enables reliable and continuous power delivery for UDs, several concerns arise when integrating WET and LoRaWAN in the SAGUIN system. First, inappropriate time allocation between WET and data transmission phases can degrade packet delivery ratio and reduce operational lifetime of UDs. Second, the LoRaWAN SF should be carefully configured to ensure reliable connectivity while minimizing transmission energy consumption.} For instance, a step increase in SF almost doubles the time-on-air (ToA) thus increases the maximum radio link distance and the probability of collisions. Third, the radio channel is fluctuated by the mobility of UAVs and LEO satellites, as well as the dynamic underground conditions caused by precipitation. \textcolor{b}{Therefore, both time allocation and SF configuration should be adaptively adjusted based on the current estimated CSI to establish a robust and efficient SAGUIN system. Accordingly, in Section~\ref{SimRes}, we jointly optimize the LoRaWAN SF configuration and the time allocation to maximize system lifetime, thereby advancing our objective of sustainable and large-scale underground monitoring in remote and post-disaster areas.}    

\subsection{Use Scenarios}
The proposed SAGUIN system powered by LoRaWAN and WET technologies promises to deliver substantial social and economic benefits across different verticals.
\begin{itemize}
    \item \textbf{Rural and Remote Areas:} The SAGUIN system promotes smart agriculture by enabling the timely collection of accurate sensor data and precise remote operations to enhance crop production on remote farms. Additionally, it provides technological assistance for real-time leakage detection and localization of underground pipelines in rural areas.
    
    \item \textbf{Urban and Suburban Regions:} For large-scale underground infrastructure monitoring in urban and suburban regions, the proposed system provides cost-effective data transmission, extends coverage in obstructed underground environments, and supports real-time data access with resilience against environmental disruptions, which is not always guaranteed by local terrestrial networks.
    
    \item \textbf{Post-disaster Areas:} In post-disaster scenarios where aboveground communication infrastructure is damaged or destroyed, the SAGUIN system enables timely geological monitoring and accurate localization of trapped individuals and essential assets within hazardous zones.
\end{itemize}

The annual deployment costs of the SAGUIN system for these scenarios are summarized in Table~\ref{tab_costs}. Specifically, remote deployments are costliest due to UAV use, off-grid solar-plus-battery beacons, and potential satellite launches; urban deployments are cheapest by leveraging the existing power grid and satellite constellations; and post-disaster cases lie in between owing to mobile power systems and UAV requirements.

\begin{table*}[!t]
\centering
\caption{Annual Deployment Costs and Power Sources for Our Proposed SAGUIN System}
\label{tab_costs}
\begin{tabular}{|>{\centering\arraybackslash}m{2.2cm}|p{4.2cm}|p{2cm}|p{1.4cm}|p{1.2cm}|p{2.8cm}|}
\hline
\textbf{Scenario} & \textbf{Equipment / Facility} & \textbf{Unit Cost} & \textbf{Quantity} & \textbf{Cost} & \textbf{Primary Power Supply} \\
\hline
\multirow{8}{*}{\textbf{Remote Area}} & NTN gateway & \$1,000 & 1 & \$1,000 & \multirow{8}{=}{Solar panel array paired with battery storage.} \\
& Remote power beacon & \$5,000 & 1 & \$5,000 & \\
& UDs & \$30 & 1,000 & \$30,000 & \\
& WET-enabled UAV & \$8,000 & 1 & \$8,000 & \\
& HAP rental cost & \$2,000/month & 12 & \$24,000 & \\
& Satellite launch cost & \$10,000 & 1 & \$10,000 & \\
& Satellite backhaul link & \$60/GB & 100 & \$6,000 & \\
&Cloud services	&\$300/month	& 12	&\$3,600 &\\
\cline{2-5}
& \multicolumn{4}{r|}{\textbf{Total Cost: \$87,600}} & \\
\hline

\multirow{7}{*}{\textbf{Urban Area}} & NTN gateway & \$1,000 & 1 & \$1,000 & \multirow{7}{=}{The existing power grid.} \\
& Urban power beacon & \$3,000 & 1 & \$3,000 & \\
& UDs & \$30 & 1,000 & \$30,000 & \\
& HAP rental cost & \$2,000/month & 12 & \$24,000 & \\
& Satellite backhaul link & \$60/GB & 100 & \$6,000 & \\
& Cloud services	&\$300/month	& 12	&\$3,600 &\\
\cline{2-5}
& \multicolumn{4}{r|}{\textbf{Total Cost: \$67,600}} & \\
\hline
\multirow{8}{*}{\textbf{Post‐disaster Area}} & NTN gateway & \$1,000 & 1 & \$1,000 & \multirow{8}{=}{Mobile diesel generators and portable solar panels with battery backup.} \\
& Movable power beacon & \$12,000 & 1 & \$12,000 & \\
& UDs & \$30 & 1,000 & \$30,000 & \\
& WET-enabled UAV & \$8,000 & 1 & \$8,000 & \\
& HAP rental cost & \$2,000/month & 12 & \$24,000 & \\
& Satellite backhaul link & \$60/GB & 100 & \$6,000 & \\
& Cloud services	&\$300/month	& 12	&\$3,600 &\\
\cline{2-5}
& \multicolumn{4}{r|}{\textbf{Total Cost: \$84,600}} & \\
\hline
\end{tabular}

\begin{minipage}{0.9\linewidth}
\vspace{1mm}
\textit{Note: All listed products are commercially available, and the costs presented above exclude labor and logistics for equipment deployment, field installation, and long-term maintenance, which may vary substantially across regions.}
\end{minipage}

\vspace{-3mm}
\end{table*}

\section{SAGUIN System over LoRaWAN and WET: Use Case and Proof of Concept}~\label{SimRes}
\vspace{-7mm}
\subsection{Simulations Parameters and Configurations}
\begin{figure*}[!t]
\centering
\includegraphics[width=6.5in]{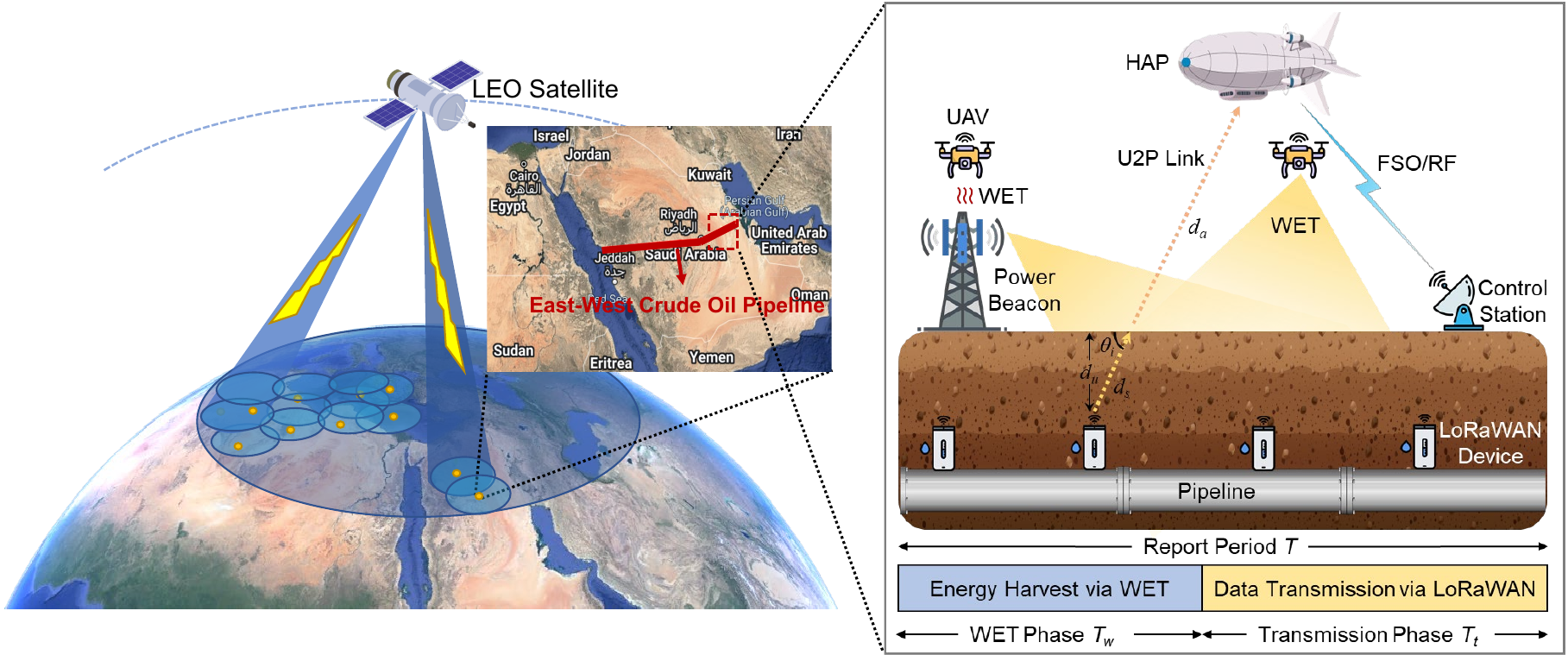}
\caption{Simulated underground pipeline monitoring scenarios in Saudi Arabia and the corresponding system model, where each UD buried at a depth of $d_u$ harvests energy from UAVs and power beacons via WET during $T_w$ and utilizes the harvested energy to transmit a 10-byte packet to the HAP during $T_t$ within a reporting period of $T$.}
\label{fig_2}
\end{figure*}

To evidence the feasibility of integrating LoRaWAN and WET to extend the operational lifetime of the SAGUIN system, we consider a realistic underground crude oil pipeline monitoring scenario in Saudi Arabia, as illustrated in Fig.~\ref{fig_2}. The corresponding key parameters of system model are summarized in Table~\ref{ParaTab}. 

Within each reporting period $T$, $N=10,000$ (i.e., 10k) UDs buried at the same depth $d_u$ harvest energy from both aboveground power beacons and UAVs through WET during $T_w$ and then leverage this harvested energy to randomly and uniformly upload a $10$-byte sensor data packet during $T_t=T-T_w$ to the network server through a LEO satellite. Note that the time allocation for the WET and data transmission phases should be carefully formulated to extend the operational lifetime of the UDs. To ensure the link reliability between the UDs and the LEO satellite, a HAP is employed as a relay to aggregate and forward all sensor data to the satellite. Considering a reliable P2S channel with no packet loss, the simulation focuses on assessing the success probability and energy efficiency of the direct U2P connectivity. Furthermore, we assume that both the WET and data transmission channels experience strong line-of-sight propagation, which implies the exclusion of RIS in this study and sets the path loss exponent $\eta = 2$.

\begin{table}[!t]
\caption{System Model Parameters} 
\centering
\begin{tabular}{m{0.26\textwidth}<{\raggedright} m{0.14\textwidth}<{\centering}}%
\toprule
\textbf{Parameters}                 & \textbf{Values}                \\  \hline
\multicolumn{2}{l}{\textbf{Operation Environment}}\\ \hline
Number of UDs ($N$)                 &10,000                            \\
Burial depth ($d_u$)                &0.6~m                            \\
VWC ($m_v$)                         &11.9\%~(\textit{in-situ})        \\
Clay ($m_c$)                        &16.86\%~(\textit{in-situ})       \\
Frame payload ($P_{rp}$)            &10~bytes                          \\
Reporting period ($T$)              &1800~s                            \\ 
Traffic pattern                     &Periodic                         \\
\hline
\multicolumn{2}{l}{\textbf{HAP Configuration}}        \\ \hline
HAP height ($H$)                    &20~km  \\
Elevation angles ($E$)              &\(30^\circ \leq E \leq 90^\circ\)\\
Coverage radius ($R$)               &35~km \\
Link distance ($d_a$)               &20$\sim$40~km                \\ \hline
\multicolumn{2}{l}{\textbf{Radio Configuration}}                        \\ \hline 
Transmit antenna gain of UDs  ($G_{tx}$)          &2.15~dBi   \\
Receiver antenna gain of gateway ($G_{rx}$)       &25~dBi     \\  
Path loss exponent ($\eta$)                       &2           \\
SIR threshold (\(\gamma\))                        &6~dB        \\
\hline
\multicolumn{2}{l}{\textbf{LoRaWAN Configuration}}                  \\  \hline
Carrier frequency ($f_c$)            &868~MHz                       \\ 
Transmit power ($P_{tx}$)            &14~dBm                         \\
Transmit current ($I_{tx}$)          &114~mA                         \\
${SF}$                               &7 $\sim$ 12                    \\
${BW}$                               &125~KHz                         \\
${CR}$                               &4/5                             \\
Frequency channels ($N_c$)           &64                              \\
Noise power (\(\sigma^2_w\))         &-117~dBm                         \\ \hline
\multicolumn{2}{l}{\textbf{Energy Model}}                  \\  \hline
Received power ($P_r$)               &0.02~W                 \\
Operating voltage of UDs ($V$)       &3.3~V                  \\
Battery capacity ($B_c$)             &3000~mAh               \\
Energy conversion efficiency ($\xi$) &0.6                   \\
\bottomrule
\label{ParaTab}
\end{tabular}
\end{table}

We adopt the modified Friis model based on the mineralogy-based soil dielectric model and set the \textit{in-situ} soil characteristics, e.g., VWC and clay percentage, to accurately capture the underground path loss~\cite{LinLoRaWUSNs}. We envision a LoRaWAN gateway deployed on a HAP that provides a spot beam covering all UDs, with an antenna gain of $25$~dBi and an evaluation angle ranging from $30^\circ$ to $90^\circ$. The system operates in the EU$868$~MHz ISM band, where all UDs are configured with the maximum allowed transmit power of $14$~dBm, an antenna gain of $2.15$~dBi, and a fixed LoRaWAN SF without adaptive data rate. Furthermore, a maximum of $64$ channels are used for uplink communication by employing a bandwidth of $125$~kHz and SF configurations ranging from SF$7$ to SF$12$. Accordingly, the ToA of a single $10$-byte uplink packet is~$T_{toa}=\{61.696, 113.152, 155.648, 370.688, 823.296, 1482.752\}$~ms for SF$7\sim12$, respectively. 

We use the success probability $P_S$ to characterize the probability of a packet being correctly received by the LoRaWAN gateway~\cite{AsadTII}. It is the product of two components: the probability of exceeding the demodulated SF-specific signal-to-noise (SNR) threshold ($D_{SNR}=\{-6, -9, -12, -15, -17.5, -20\}$~dB for SF$7\sim12$, respectively), denoted as $P_{SNR}$, and the probability of the packet interference not blocking the reception of a packet accounting for the capture effect, denoted as $P_{SIR}$. To leverage the capture effect, the signal-to-interference ratio (SIR) of the target packet should be above a certain threshold $\gamma=6$~dB. The energy per packet (EPP) represents the average amount of energy consumed by a UD to deliver a packet to the gateway, which is calculated as the ratio of energy consumed per packet to the success probability~\cite{LinMag}. We employ the average battery lifetime across all UDs to reveal the system lifetime, where the battery lifetime of each UD is determined by dividing the total available energy (which comprises both harvested energy and initial battery capacity) by the energy consumed by the UD. For this calculation, we adopt the complete LoRaWAN Class A energy consumption model in~\cite{LoRaEnergy}, assume a battery capacity of $3000$~mAh, and consider a harvested power of $0.02$~W with a conversion efficiency of $0.6$~\cite{LinMag}. Note that the harvested energy level depends on the transmit power of the power sources, underground conditions, and harvesting efficiency, and should be empirically determined through field measurements, which is left for our future work.

\subsection{Selected Numerical Results}
\begin{figure}[!t]
\centering
\includegraphics[width=3in]{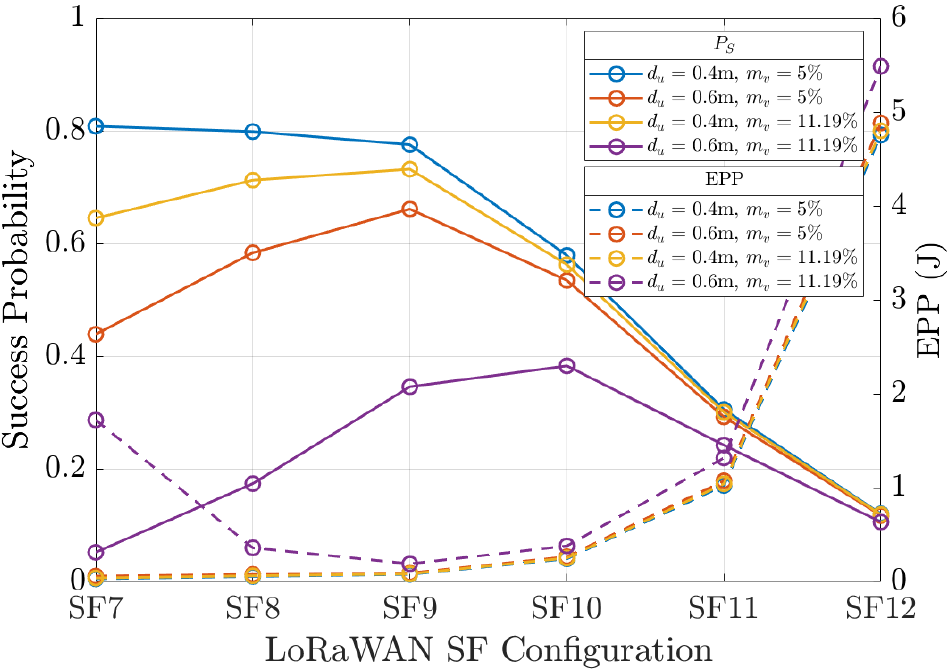}
\caption{Average success probability $P_S$ and EPP across all UDs for various burial depths and VWC values as a function of LoRaWAN SF configurations.}
\label{fig_3}
\end{figure}

First, we analyze the feasibility of direct U2P connectivity. Fig.~\ref{fig_3} illustrates the effects of UDs' burial depth, VWC of soil, and LoRaWAN SF configurations on the average success probability $P_S$ and EPP. As expected, increasing either burial depth or VWC results in a lower $P_S$ and a higher EPP due to larger attenuation in underground soil. Under favorable conditions (i.e., $d_u=0.4$~m and $m_v=5\%$), SF$7$ outperforms all other SF configurations in terms of both success probability and EPP. This is because SF$7$ has the shortest ToA, which reduces both collision probability and transmission energy consumption, especially in shallow-depth and low-VWC conditions where all SFs can maintain sufficient link budgets. In more challenging \textit{in-situ} conditions (i.e., $d_u=0.6$~m and $m_v=11.19\%$), SF10 achieves the highest success probability among all SFs, as it offers more reliable connectivity with reduced interference. However, SF$9$ shows a lower EPP than SF$10$. Although SF$10$ provides higher $P_S$, its longer ToA ($370.688$~ms) leads to higher transmission energy consumption, whereas SF9, with a shorter ToA of $155.648$~ms, consumes less energy despite a slightly lower $P_S$.

\begin{figure}[!t]
\centering
\subfloat[]{
    \includegraphics[width=3in]{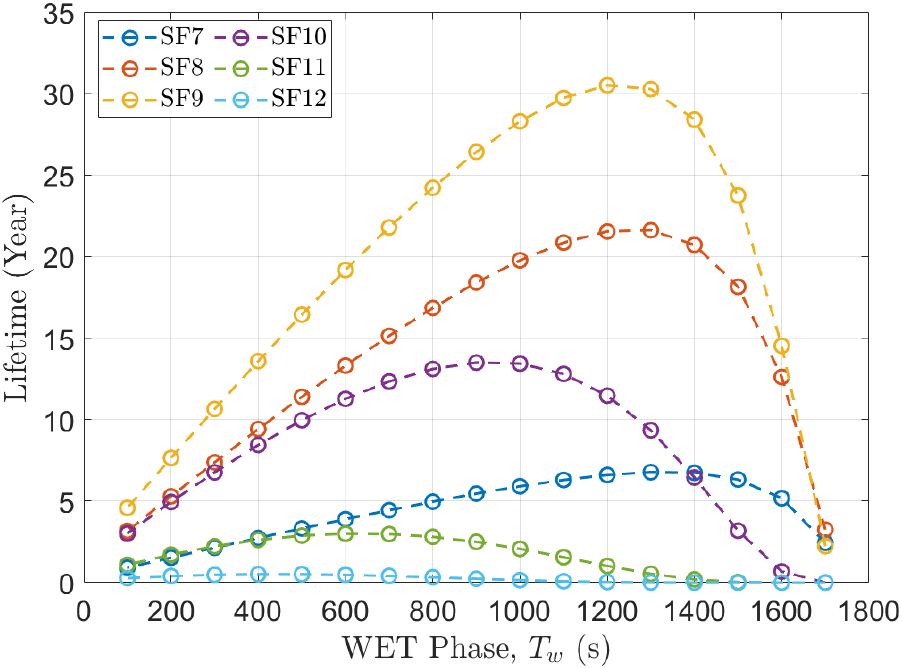}%
    \label{fig_4}
}\hfil
\subfloat[]{
    \includegraphics[width=3in]{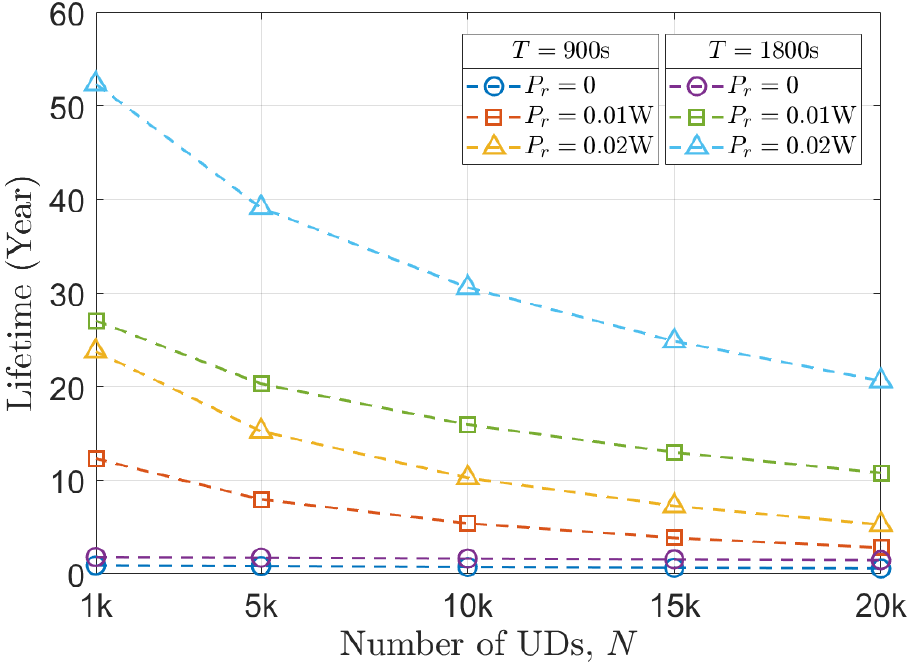}%
    \label{fig_5}
}
\caption{(a) Average lifetime over all UDs under different LoRaWAN SF configurations as a function of WET phase duration $T_w$; (b) Average lifetime of all UDs for two reporting periods versus the number of UDs at different harvested energy levels, considering the derived optimal SF configuration and time allocation, where $P_r = 0$ indicates that the UDs are powered solely by their pre-equipped batteries without WET.}
\end{figure}

Then, we study the impact of SF configurations and time allocation between the WET and data transmission phases on system lifetime. Within $T=1800$~s, each UD harvests energy during the allocated WET duration $T_w$ and uploads one packet during the subsequent data transmission period $T_t$. As shown in Fig.~\ref{fig_4}, the system lifetime initially increases and then decreases as more time is allocated to the WET phase across all SF configurations. Meanwhile, each SF configuration exhibits a unique optimal WET duration. For instance, SF$9$ achieves the longest system lifetime of $30.5$ years when $T_w=1200$~s. Although a longer WET duration allows UDs to harvest more energy, it also reduces the available time for data transmission, thereby increasing the collision probability and the EPP. Therefore, selecting an appropriate SF configuration along with its corresponding optimal WET duration is crucial for maximizing system lifetime.


Finally, we present the relationship among the reporting period $T$, harvested power $P_r$, number of UDs $N$, and system lifetime under the optimal SF configuration and its corresponding time allocation. Note that $P_r=0$ implies that the UDs rely solely on their pre-equipped batteries without any harvested energy. To this end, we employ bounded numerical optimization to determine the optimal WET duration for each SF configuration and subsequently identify the one that maximizes system lifetime. As illustrated in Fig.~\ref{fig_5}, system lifetime decreases with an increasing number of UDs due to the higher probability of collisions. For instance, under $T_w=1800$~s and $P_r=0.02$~W, the system lifetime drops from $52.4$ years to $20.6$ years as the number of UDs increases from $1$k to $20$k. The results also reveal that the system lifetime under $T = 900$~s is significantly shorter than that under $T = 1800$~s, due to two main factors: (i) increased total energy consumption caused by more uplink packets, and (ii) higher collision probability resulting from the shorter reporting period. Furthermore, with $T_w = 1800$~s and $N = 20$k, the system lifetime at $P_r = 0.02$~W is approximately $30$ times longer than in the case without WET (i.e., $P_r = 0$). Note that the operational lifetime of UDs in real deployments may be lower than these simulated values due to reduced harvested power under practical conditions and additional constraints from the lithium-battery's cycle life, hardware degradation, corrosion, and component aging in harsh underground soils.



\section{Open Challenges and Research Directions}
The above quantitative analysis demonstrates the feasibility of integrating LoRaWAN and WET within the proposed SAGUIN system to extend the operational lifetime of UDs. Note that the proposed SAGUIN architecture can be generalized to other underground scenarios, such as smart agriculture and post-disaster rescue operations. However, several potential challenges still need to be carefully addressed when integrating LoRaWAN and WET for practical SAGUIN deployment.
\begin{itemize}
    \item \textbf{Channel Modeling:} To accurately evaluate the practical performance of the proposed SAGUIN system, the channel model between UDs and various NTN platforms shall be carefully characterized by accounting for multi-layer signal absorption in the underground portion, refraction loss in the soil-air interface, and empirical air path loss models across different elevation angles.

    \item \textbf{CSI Acquisition:} CSI estimation is critical for optimizing time allocation, SF configuration, and WET beamforming within the SAGUIN system. However, acquiring reliable and accurate CSI remains a significant challenge in subterranean mMTC scenarios due to the harsh underground environment and the heterogeneous and time-varying propagation conditions inherent to SAGUIN. This underscores the need for low-complexity and energy-efficient CSI estimation protocols that reduce the number of training pilots and minimize energy consumption among UDs. Moreover, the accuracy of CSI can be further enhanced by leveraging deep learning-based CSI estimation techniques.
    
    \item \textbf{Coordination Mechanisms:} As the SAGUIN system is a heterogeneous multi-layer network, multiple transmission paths exist to meet diverse service requirements. Accordingly, an adaptive routing strategy should be developed to ensure reliable and energy-efficient data transmission, considering the properties of the underground, terrestrial, aerial, and space segments. Furthermore, since NTN platforms follow distinct and region-dependent frequency plans and regulations, it is worthwhile to develop platform-aware frequency-switching protocols and regulatory frameworks to enable efficient handovers among these platforms.
    

    \item \textbf{Scalability and Interference Management:} The network capacity and collision resilience of LoRa are inherently limited by its ALOHA-like medium access protocol. To address this scalability challenge, the LoRaWAN protocol has recently incorporated LoRa-frequency hopping spread spectrum (LR-FHSS), which enhances reliability and interference robustness in uplink mMTC through improving spectral efficiency and introducing redundancy in coding and physical headers. Notably, LR-FHSS can co-exist with LoRa modulation, allowing UDs to adaptively switch between modulation schemes for reliable and energy-efficient communication in dynamic underground environments. 
    
    \item \textbf{Efficient WET Operation:} To improve WET efficiency from power sources to UDs, several promising research directions can be explored, including CSI-free WET schemes, efficient UAV trajectory planning, and the RIS deployment. In particular, the phase shifts and positioning of the RIS, time allocation, and LoRaWAN SF configuration shall be jointly optimized to enhance channel gains. Furthermore, sustainable subterranean mMTC can be facilitated by powering energy transmitters with ambient energy sources, which in turn motivates further investigations into energy scheduling and cooperation protocols.
\end{itemize}

\section{Conclusion}
To extend the emerging SAGIN paradigm into the subterranean domain, in this article, we proposed the SAGUIN architecture, which integrates LoRaWAN and WET technologies to enhance both communication and energy capabilities. After introducing the technical background, system architecture, and implementation strategy for SAGUIN, we assessed its feasibility and performance in a realistic underground pipeline monitoring scenario. Our numerical results demonstrated that the proposed SAGUIN system, enabled by the integration of LoRaWAN and WET, could achieve sustainable subterranean mMTC. Moreover, the system lifetime was further extended by jointly optimizing time allocation and LoRaWAN SF configurations. Nevertheless, its performance was inevitably affected by the number of UDs, underground parameters, reporting periods, and harvested energy levels. Finally, we identified key future research directions to motivate deeper theoretical and experimental exploration for this novel concept.
\bibliographystyle{IEEEtran}
\bibliography{ref}



\end{document}